# A Transactive Retail Market Mechanism for Active Distribution Network Integrated with Large-scale Distributed Energy Resources

Chunyi Huang, *Student Member, IEEE,* Chengmin Wang, Mingzhi Zhang, *Student Member, IEEE,* Ning Xie, Yong Wang, *Senior Member, IEEE*

*Abstract*—The burgeoning integration of distributed energy resources (DER) poses new challenges for the economic and safe operation of the electricity system. The current distribution-side policy is largely based on mandatory regulations and incentives, rather than the design of a competitive market mechanism to arouse DERs to freely compete in the retail market. To address this issue, we proposed a transactive retail market mechanism to attract the active participation of profit-driven DER retailers in a deregulated way. Considering the limited competitive property of the retail market, a bi-level DSO-dominated framework is constructed to simulate the virtual game between distribution system operator (DSO) and DER retailers. Specifically, the flexible interval pricing of retailers is modeled as a series of binary revenue constraints to influence the DSO's centralized economic dispatch decisions, where the time-varying distribution locational marginal price is adopted to settle energy activities at different locations over the time horizon. Due to the uncertain power flow path during the coordinated decision-making process, we proposed a general undirected second-order cone-based AC radial power flow model and provided sufficient conditions to ensure its exactness. Also, we applied a series of approximation and relaxation techniques to transform the bi-level mixed-integer quadratic framework with highly discrete induced domains into a solvable mixed-integer semidefinite programming problem. It is demonstrated in the case study that the proposed mechanism can not only improve market efficiency but also eliminate market power and the resulting market failures.

## Nomenclature

*A. Indices*
$i$ — Node index.
$j$ — Adjacent node index with bus i.
$t$ — Time index.

*B. Sets*
$\Psi_N, \Psi_{\{1\}}$ — Number of nodes and the substation node.
$\Psi_w, \Psi_v, \Psi_c$ — Number of wind, PV retailers and shunt banks.
$\Psi_f$ — Number of feeders.
$\Omega_i^+, \Omega_i^-$ — Number of lines flowing in and out of i.
T — Number of time slots.

*C. Parameters*
$R_i, \theta_i$ — Cumulative and unit revenue thresholds.
$\overline{c}_{it}, \underline{c}_{it}$ — Boundary of pricing interval.
$LMP_t$ — Locational marginal price of power injection.
$c^{tr}$ — Unit carbon emission cost.
$r^{gt}$ — Rate of carbon conversion.
$r_{ij}, x_{ij}$ — Impedance of line i-j.

*D. Variables*
$c_{it}$ — Pricing bids of retailer i at time slot t.
$u_i$ — Participation status of retailer i in the market.
$\tau_{it}^p$ — DLMP of bus i at time slot t.
$z_{it}^g, z_{it}^g$ — Utilization label of retailer i at time slot t.
$P_t^{tr}, Q_t^{tr}$ — Active and reactive power injections.
$P_{it}^w, P_{it}^v, Q_{it}^w, Q_{it}^v$ — Active and reactive power of WG and VG.
$Q_{it}^c$ — Reactive power compensation quantity.
$v_{it}$ — Squared voltage amplitude of bus i at slot t.
$z_{ijt}^{+,-}, \tilde{z}_{ijt}^{+,-}$ — Integrity and relaxed power flow direction.
$P_{ijt}^{+,-}, Q_{ijt}^{+,-}, l_{ijt}^{+,-}$ — Forward and reverse active and reactive power, and squared current through the line.

## I. Introduction

THE claims of energy transition promote the extensive adoption of distributed energy resources (DER) in the distribution system, especially the residential rooftop photovoltaic systems owned by end-users [1]. Due to the lack of advanced communication and control systems, existing applications of geographically dispersed DERs mainly focus on local and even domiciliary daytime supply, yet neglecting the potential of using large-scale DER integration to flatten peak demands of the entire network.

In general, the local utility will dispatch controllable equipment like regulators to maintain secure operation when hosting medium- or small-scale DERs [2]. With the increasing penetration ratio of DERs, it may result in non-deferrable investments to ameliorate the flexibility requirements under uncertain operating conditions. Besides, from the perspective of market operation, the regulatory tariff is heavily subsidized by governments to encourage the installation and penetration of DERs [3], which increases the actual consumption cost of clean energy and makes uneconomical energy evolvement. According to the techno-economic analysis given in [4], the viability of residential DER completely depends on contractual compensations. As incentives are gradually diminishing in the process of market deregulation, designing an effective and competitive retail market mechanism that enables the active involvement of stakeholders is an imperative step.

In the deregulated retail market, the local utility previously monopolized the distribution system will serve as a neutral distribution system operator (DSO) to monitor business activities among entities [5]-[6]. A conceptual stakeholder, the aggregator/agent, has been introduced in multi-agent-based works to manage the energy-related behavior of dispersed end-users [7]-[8]. Unlike the traditional centralized clearing where bids and offers are collected offline, the interoperability among entities is considered in these studies, where a game theory-based hierarchical framework is built to reflect differentiated decision preferences [9]-[11]. In a multi-lateral retail market proposed in [10], the retailer at the lower level will adjust its pricing strategy according to the scheduling scheme formulated by DSO, thereby reversely affecting the economic operation.

Inspired by growing initiatives on the transactive paradigm, the time-varying clearing price is exploited to embody the price elasticity of players when formulating energy management schemes [12]-[13] and designing trading strategies [14]-[15]. The multi-level framework is extensively adopted to exchange price signals and demands among stakeholders at different levels. As elaborated in [12], a tri-layer model is designed to simulate the hierarchical dispatching of a micro-grid community in inner-micro-grid, inner-aggregator, and inter-aggregator levels, where the spot price of micro-grid is adopted to stimulate the energy trading within the community. This price is obtained by solving the unbalanced power function composed of price-quantity functions of price-takers including DES aggregators, energy storage systems (ESS), and flexible loads. In addition to the clearing price based on the market theory, the profit-driven pricing is also exploited. In [14], a bi-level model is constructed to design the retailer's pricing strategy considering demand elasticity. Specifically, the distribution-level retail price will be released by the upper-layer retailer to maximize profit, and the demand for air-conditioning is then adjusted by lower-level agents to compromise the consumption satisfaction and costs.

Although the unified clearing price can motivate the responsiveness of stakeholders in a decentralized way, this price dependent on demand and supply relationship exposes its drawbacks on neglecting physical operational models and homogenizing nodal cost. Considering the growing number of market participants, the distribution locational marginal price (DLMP) acts as a promising tool to precisely value the energy transactions within the distribution system [16]. Given the nonlinearity of AC power flow, the calculation of DLMP is categorized into the linearization-based [17]-[18] and second-order cone relaxation-based (SOC) methods [19]-[20]. The former and its variants use linearization methods to estimate and calibrate network losses and voltage drops on a Dist-Flow model, which has to calculate repeatedly to find the coefficients in formulas. The latter methodology outperforms in computing performance due to the convex property, whilst the exactness of its solution is limited by a series of strict sufficient conditions. Notably, these conditions put ex-ante restrictions on the impedance distribution and predefined radial power flow path [21], unbinding status of nodal voltage [22], and power injection [23], which undoubtedly limits its applicability to realistic radial circuits hosting large-capacity DERs.

In theory, due to geographical delivery capabilities and jurisdictional limitations, the hypothesis of perfect competition may no longer apply to the deregulated retail market, so the price bidding of DER aggregators who act as price-makers will directly influence the market settlement. Although for clusters such as microgrids [24] or multi-carrier energy systems [25], market settlement estimates are considered in the upper-level supplementary bidding strategy design of the bi-level model, this rough speculation of complete information may run counter to reality, leading to loss of profits or cuts in clean energy in the final clearing. It also exposes the inherent obstacles of traditional sequential market mechanisms in forming profitable pricing strategies to attract the active participation of DERs and in maintaining energy penetration while removing regulation.

To address these barriers, we presented a transactive retail market mechanism by simultaneously formulating the price bidding of retailers and the centralized market-clearing in a coordinated fashion. Given that the compound bidding not only increases the access qualification of end-users with small-capacity DERs, but also limits possible market business models, we studied the pricing bidding strategy of DER clusters in the retail market in the form of homogeneous nodal aggregators, namely DER retailer hereafter. In contrast to forming an on-site game between these entities, we decoupled the interval pricing of the DER retailer from its output and further introduced binary revenue constraints to replace their profitability considerations to form a virtual game in an interactive bi-level framework [26]. Also, to overcome the bottleneck of DLMP formulations which is based on the classic SOC power flow model on characterizing uncertain bi-directional power flow [19], we developed a novel undirected SOC-based AC power flow model along with its sufficient conditions on maintaining exactness for radial distribution circuits and applied this model to calculate spatial-temporal differentiated DLMP which is employed it to settle transactions in the proposed market mechanism. By transforming the bi-level framework into a centralized decision made by DSO, the analogous equilibrium solution regarding the pricing bids of DER retailers and the whole clearance is formed, thus eliminating market failures caused by the contradiction between enhancing clean energy penetration and satisfying the capital recovery of DERs. The main contributions are summarized as:

1) A bi-level transactive retail market mechanism is proposed for the active distribution network integrated with large-scale DERs. A day-ahead market settlement scheme and decoupled pricing bids of DER retailers are obtained simultaneously to enhance market efficiency without direct interventions.

2) A novel undirected second-order cone relaxation-based AC power flow model for radial circuits is presented. This model significantly expands the application scenario of the classic formulation, whilst closed-form proof is also provided to prove the exactness of the derived solution under mild and realistic sufficient conditions.

3) A mixed-integer semidefinite programming (MISDP) is constructed to intensively solve the proposed bi-level market framework by applying a series of approximation and relaxation techniques. It is validated that the calculation accuracy of this reformulation is acceptable in the process of coordinated decision-making.

The rest of this paper is organized as follows. The details of the proposed transactive market mechanism are presented in section II. The reformulation along with the relaxation techniques of the bi-level framework are given in section III. The case study is elaborated on section IV. The conclusion and prospective works are given in section V.

## II. Transactive Retail Market Mechanism

### A. Entities in the transactive retail market

The entities in the retail market are divided into DSO and stakeholders that include load aggregators and DER retailers. For simplicity, the electricity generation and usage are proxied by separate players in this manuscript. The DER retailer, specifically the wind and photovoltaic (PV) aggregators contracting adjacent renewable generation units, acts as a seller in designing pricing bids and managing operation scheduling. The responsibility of the load aggregator is to submit offers of the cumulative demand of nodal consumers as a price-taker. According to the initiative advocated in the past decade, the role of DSO evolves into a neutral facilitator to supervise systematic and market operations [5]. Its overriding principle in the proposed market mechanism is to organize the bids and offers and make the centralized market clearing.

### B. DSO-dominated Transactive Retail Market Settlement

A transactive retail market mechanism is proposed to preserve the profitability of DER retailers without rewarding subsidies. Given that the pricing of the DER retailer is critical to the settlement scheme of a limited competitive market, a bi-level DSO-dominated framework is established in Fig.1 to coordinate the dynamic pricing and the market settlement.

Compared with the traditional sequential bidding and market clearing, it is observed the interplay between entities relies on a bi-directional pricing and revenue exchange, where the counter-effect of market-clearing on retailer pricing is firstly considered. To avoid market power caused by conspiracy to manipulate clearing prices, this framework is independently resolved by DSO. The stakeholders are obligatory to submit bids and offers before the permittable timepoint in a wait-and-see fashion, whereas DSO will collect and detect all information, and execute the transactive market settlement impartially.

It is worth noting that profit-seeking decisions of retailers are not modeled as separate models, but as a series of binary revenue constraints submitted by retailers to DSO to influence the centralized decision-making process, detailed as (1)-(4).

$$(\tau_i^\mathrm{p})^T P_i^\mathrm{g} \geq \mathrm{R}_i u_i + (\theta_i)^T P_i^\mathrm{g}, \forall i \in \Psi_w \cup \Psi_v. \quad (1)$$

$$\underline{c}_{it} \leq c_{it} \leq \overline{c}_{it}, \quad (2)$$

$$u_i \geq \sum_{t \in \mathrm{T}} z_{it}^\mathrm{g}, \quad (3)$$

$$u_i \leq z_{it}^\mathrm{g}, \forall i \in \Psi_w \cup \Psi_v, t \in \mathrm{T}. \quad (4)$$

The profitability of retailers is specifically restricted by a predetermined bidding parameter set $(\mathrm{R}_i, \theta_i, \overline{c}_{it}, \underline{c}_{it})$, which contains the cumulative revenue threshold, the minimal profitable unit price, and the upper and lower limits of the pricing interval. Equation (1) reflects the inter-temporal revenue considerations of retailers, where $P_{it}^\mathrm{g}$ is the unified dispatching output, and $\tau_{it}^\mathrm{p}$ denotes the DLMP derived from the Lagrange multiplier of active power balance constraint at time slot t. $u_i$ is a binary variable indicating the participation status of the retailer, constrained by the utilization label $z_{it}^\mathrm{g}$ in (3)-(4). If $u_i$ equals to 1, then the revenue constraint will take effect. As in (2), the pricing bids $c_{it}$ fluctuate within a series of intervals decoupled from $P_{it}^\mathrm{g}$, but there is no doubt the variation of $c_{it}$ will affect $P_{it}^\mathrm{g}$ and $\tau_{it}^\mathrm{p}$ under the economic dispatching pattern. It is worth mentioning that the involvement of binary revenue constraint only attracts competitiveness of DER retailers to submit flexible pricing bids in the market, rather than sacrificing the fairness of DSO to consume more clean energy.

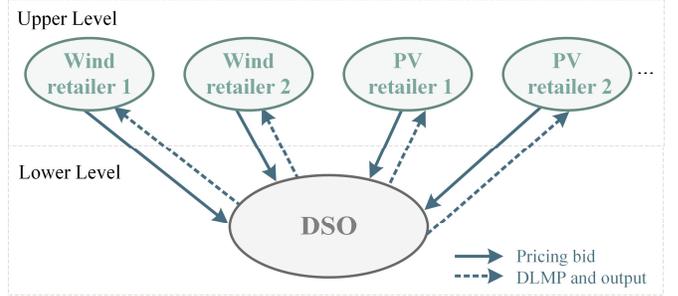

Fig. 1. The interplay between DSO and DER retailers in the transactive retail market mechanism

The mathematical formulation of this bi-level transactive market framework is listed as (5)-(28). Among them, equations (5)-(7) consist of the upper-level model with the decision variable set (6), which decides the pricing bids and participation status of retailers subjecting to income considerations. The remaining equations (8)-(28) constitute the lower-level model, which formulates the day-ahead scheduling plan to minimize the intraday generation costs composing of pricing bids.

The upper-level objective in (5) is to maximize the social welfare consisting of three parts. The first term is the security and economic value of the systematic operation, defined by the generation cost reduction in part A of (5) and (8). The second component is the carbon emission cost savings denoted as part B. Lastly, the third part refers to the generation cost savings to quantify the efficiency of the resource allocation, which is defined in C to minimize the price deviation of DER retailers between the final scheme and initial intervals. The smaller the part, the more economical the penetration of clean energy.

$$\max_Y SW = -\underbrace{\sum_{t \in \mathrm{T}}\left(\sum_{i \in \Psi_w \cup \Psi_v} c_{it} P_{it} + \mathrm{LMP}_t \cdot P_t^\mathrm{tr}\right)}_{\mathcal{A}}$$
$$-\underbrace{\sum_{t \in \mathrm{T}} c^\mathrm{tr} r^\mathrm{gt} P_t^\mathrm{tr}}_{\mathcal{B}} - \underbrace{\sum_{t \in \mathrm{T}} \sum_{i \in \Psi_w \cup \Psi_v} (c_{it} - \underline{c}_{it}) P_{it}}_{\mathcal{C}}$$
$$\quad (5)$$

$$subject\ to. \quad Y = \{c_{it}, u_i\} \quad (6)$$

$$(1)-(4) \quad (7)$$

$$\min_X \mathcal{A} \quad (8)$$

$$subject\ to. X = \begin{Bmatrix} P_{it}^\mathrm{w}, P_{it}^\mathrm{v}, P_t^\mathrm{tr}, Q_{it}^\mathrm{w}, Q_{it}^\mathrm{v}, Q_t^\mathrm{tr}, Q_{it}^\mathrm{c}, \\ P_{ijt}^{+,-}, Q_{ijt}^{+,-}, l_{ijt}^{+,-}, v_{it}, z_{ijt}^{+,-}, z_{it}^\mathrm{g} \end{Bmatrix} \quad (9)$$

$$P_t^\mathrm{tr} + P_{it}^\mathrm{g} + \sum_{j \in \Omega_i^+}(P_{ijt} - r_{ij} l_{ijt}^+) \quad (10)$$

$$Q_t^{\mathrm{tr}} + Q_{it}^{\mathrm{g}} + Q_{it}^{\mathrm{c}} + \sum_{j\in\Omega_i^-}(r_{ij}l_{ijt}^- - P_{ijt}) = P_{it}^{\mathrm{L}} : \tau_{it}^{\mathrm{p}},$$

$$Q_t^{\mathrm{tr}} + Q_{it}^{\mathrm{g}} + Q_{it}^{\mathrm{c}} + \sum_{j\in\Omega_i^+}(Q_{ijt} - x_{ij}l_{ijt}^+)$$
$$+ \sum_{j\in\Omega_i^-}(x_{ij}l_{ijt}^- - Q_{ijt}) = Q_{it}^{\mathrm{L}} : \tau_{it}^{\mathrm{q}}, \quad (11)$$
$$\forall i \in \Psi_{\mathrm{N}}, t \in \mathrm{T}.$$

$$v_{it} - v_{jt} = 2(r_{ij}P_{ijt} + x_{ij}Q_{ijt}) - (r_{ij}^2 + x_{ij}^2)l_{ijt}$$
$$: \omega_{ijt}, \forall ij \in \Psi_f, t \in \mathrm{T}. \quad (12)$$

$$\left| l_{ijt}^+ + v_{it} ; 2P_{ijt}^+ ; 2Q_{ijt}^+ ; l_{ijt}^+ - v_{it} \right| \preccurlyeq 0$$
$$: \left| d_{ijt}^+ ; d_{ijt}^{\mathrm{A}+} ; d_{ijt}^{\mathrm{B}+} ; d_{ijt}^{\mathrm{C}+} \right|, \quad (13)$$

$$\left| -l_{ijt}^- + v_{jt} ; -2P_{ijt}^- ; -2Q_{ijt}^- ; -l_{ijt}^- - v_{jt} \right| \preccurlyeq 0$$
$$: \left| d_{ijt}^- ; d_{ijt}^{\mathrm{A}-} ; d_{ijt}^{\mathrm{B}-} ; d_{ijt}^{\mathrm{C}-} \right|, \quad (14)$$

$$0 \le P_{ijt}^+ \le z_{ijt}^+ \bar{P}_{ij} : \lambda_{ijt}^{\mathrm{p}+} \ge 0, \varphi_{ijt}^{\mathrm{p}+} \le 0, \quad (15)$$

$$-z_{ijt}^+ \bar{Q}_{ij} \le Q_{ijt}^+ \le z_{ijt}^+ \bar{Q}_{ij} : \lambda_{ijt}^{\mathrm{q}+} \ge 0, \varphi_{ijt}^{\mathrm{q}+} \le 0, \quad (16)$$

$$0 \le l_{ijt}^+ \le z_{ijt}^+ \bar{L}_{ij} : \lambda_{ijt}^{\mathrm{f}+} \ge 0, \varphi_{ijt}^{\mathrm{f}+} \le 0, \quad (17)$$

$$-z_{ijt}^- \bar{P}_{ij} \le P_{ijt}^- \le 0 : \lambda_{ijt}^{\mathrm{p}-} \ge 0, \varphi_{ijt}^{\mathrm{p}-} \le 0, \quad (18)$$

$$-z_{ijt}^- \bar{Q}_{ij} \le Q_{ijt}^- \le z_{ijt}^- \bar{Q}_{ij} : \lambda_{ijt}^{\mathrm{q}-} \ge 0, \varphi_{ijt}^{\mathrm{q}-} \le 0, \quad (19)$$

$$-z_{ijt}^- \bar{L}_{ij} \le l_{ijt}^- \le 0 : \lambda_{ijt}^{\mathrm{f}-} \ge 0, \varphi_{ijt}^{\mathrm{f}-} \le 0, \quad (20)$$

$$z_{ijt}^+ + z_{ijt}^- = 1, z_{ijt}^+, z_{ijt}^- \in \{0,1\}, \forall ij \in \Psi_f, t \in \mathrm{T}. \quad (21)$$

$$\underline{v}_{it} \le v_{it} \le \bar{v}_{it} : \lambda_{it}^{\mathrm{v}} \ge 0, \varphi_{it}^{\mathrm{v}} \le 0, \forall i \in \Psi_{\mathrm{N}\setminus\{1\}}, \quad (22)$$

$$v_{it} = 1 : \omega_t^{\mathrm{bn}}, \forall i \in \Psi_{\{1\}}, t \in \mathrm{T}. \quad (23)$$

$$0 \le P_t^{\mathrm{tr}} \le \bar{P}_t^{\mathrm{tr}} : \lambda_{it}^{\mathrm{pj}} \ge 0, \varphi_{it}^{\mathrm{pj}} \le 0, \quad (24)$$

$$0 \le Q_{it}^{\mathrm{tr}} \le \bar{Q}_{it}^{\mathrm{tr}} : \lambda_{it}^{\mathrm{qj}} \ge 0, \varphi_{it}^{\mathrm{qj}} \le 0, \forall i \in \Psi_{\{1\}}, \quad (25)$$

$$0 \le P_{it}^{\mathrm{g}} \le z_{it}^{\mathrm{g}} \bar{P}_{it}^{\mathrm{g}} : \lambda_{it}^{\mathrm{pg}} \ge 0, \varphi_{it}^{\mathrm{pg}} \le 0, z_{it}^{g} \in \{0,1\}, \quad (26)$$

$$0 \le Q_{it}^{\mathrm{g}} \le z_{it}^{\mathrm{g}} \bar{Q}_{it}^{\mathrm{g}} : \lambda_{it}^{\mathrm{qg}} \ge 0, \varphi_{it}^{\mathrm{qg}} \le 0, \forall i \in \Psi_w \cup \Psi_v, \quad (27)$$

$$0 \le Q_{it}^{\mathrm{c}} \le \bar{Q}_{it}^{\mathrm{c}} : \lambda_{it}^{\mathrm{c}} \ge 0, \varphi_{it}^{\mathrm{c}} \le 0, \forall i \in \Psi_c, t \in \mathrm{T}. \quad (28)$$

It is noted that due to the time-varying nature of bidding prices, the power flow directions of the whole system cannot be postulated in advance. Thus, an undirected SOC-based AC power flow model is utilized to describe the uncertain power flow of the radial circuit. Compared with the classic SOC model discussed in [21], the proposed undirected model adapts to uncertain power flow scenarios and maintains accuracy under mild conditions. Limited by the space, the exactness proof of the solution obtained by this model is given in Appendix A [30]. Equations (10)-(23) constitute this novel power flow model, where binary variable pairs $z_{ijt}^{+,-}$ are introduced to restrict the uniqueness of power flow direction at each time granularity. Since the power factor of end-users is close to 1, it is assumed that the active power flow direction is consistent with that of the line current. For brevity, the state variables including $P_{ijt}, Q_{ijt}, l_{ijt}$ are extracted into algebraic summations of the forward and reverse solutions, e.g. $P_{ijt} = P_{ijt}^+ + P_{ijt}^-$. As illustrated in [22], $v_{it}, l_{ijt}$ are the squared values of nodal voltage and current respectively, while $l_{ijt}^-$ represents the opposite value of squared line current. The symbol $\preccurlyeq$ in (13)-(14) indicates the rotated second-order cone calculator. Equations (24)-(28) limit the available capacity of substation injection, DER retailers, and reactive power compensation, whereas the binary variable $z_{it}^{\mathrm{g}}$ in (3)-(4) and (26)-(27) reflects the utilization status of retailer $i$ in timeslot $t$. The variables $\omega, \lambda, \varphi$ at the end of constraints are their Lagrange multipliers.

## III. APPROXIMATIONS AND RELAXATIONS OF THE BI-LEVEL MARKET FRAMEWORK

Since the inducible domain of the proposed bi-level market framework is highly non-continuous in the solution space, we reformulated this structure into a solvable MISDP by using the primal-dual relaxation and approximation techniques.

### A. Strong duality relaxation for economic dispatching model

Substituting the lower-level linear programming to equivalent Karush-Kuhn-Tucker conditions is a common way to split a bi-level structure and obtain the primal and dual solutions simultaneously. Nevertheless, performing this conversion directly in the proposed lower-level mixed-integer second-order cone programming (MISOCP) will introduce complementary trilinear products with an unavoidable duality gap. Inspired by the primal-dual formulation for a mixed-integer linear model in [27], we relaxed the integrity of MISOCP and recovered the strong duality of this convex SOCP to eliminate complementary slackness conditions.

Firstly, binary variables $z_{ijt}^{+,-}, z_{it}^{\mathrm{g}}$ in (21) and (26)-(27) are relaxed to be continuous as $\tilde{z}_{ijt}^{+,-}, \tilde{z}_{it}^{\mathrm{g}}$ constrained by (29)-(30).

$$0 \le \tilde{z}_{ijt}^+ \le 1 : \lambda_{it}^{\mathrm{z}+} \ge 0, \varphi_{it}^{\mathrm{z}+} \le 0,$$
$$0 \le \tilde{z}_{ijt}^- \le 1 : \lambda_{it}^{\mathrm{z}-} \ge 0, \varphi_{it}^{\mathrm{z}-} \le 0. \quad (29)$$
$$\tilde{z}_{ijt}^+ + \tilde{z}_{ijt}^- = 1 : \omega_{it}^{\mathrm{z}}, \forall ij \in \Psi_f, t \in \mathrm{T},$$

$$0 \le \tilde{z}_{it}^{\mathrm{g}} \le 1 : \lambda_{it}^{\mathrm{u}} \ge 0, \varphi_{it}^{\mathrm{u}} \le 0, \forall i \in \Psi_w \cup \Psi_v, t \in \mathrm{T}. \quad (30)$$

Due to space limitations, the equivalent dual model of the SOCP is provided in Appendix B [30]. Since the slater conditions are met, the strong duality property is also held, illustrating the duality gap reflecting the difference between the primal and dual objectives, denoted as $\mathcal{A} - \mathcal{D}$, will equal to zero. Based on the proof in Appendix C [30], minimizing this defined duality gap is theoretically equivalent to the complementary slackness conditions. Therefore, the Karush-Kuhn-Tucker conditions of this SOCP are approximated by a new model that minimizes the duality gap restricted by the primary constraints in (9)-(20) and (22)-(30) and dual constraints in Appendix B.

However, since the relaxation in (29)-(30) may result in decimal solutions, binary stints in (21) and (26)-(27) are then supplemented to restore the integrity of relaxed variables. Therefore, a reformulated MISOCP is established as below. It is noted that due to the non-smoothness of the feasible domain, the optimal solution will be close to but not zero.

$$\min \mathcal{A} - \mathcal{D} \quad (31)$$

$$\begin{aligned} subject\ to.\ &primal\ constraints: (9)-(30) \\ &dual\ constraints\ in\ Appendix\ B \end{aligned} \quad (32)$$

To further simplify the model structure, (31) is approximated by an inequality constraint as (33) restricted by a small positive constant $\varepsilon_\Delta$. Thus, the lower-level MISOCP model is transformed into a constraint set including (32)-(33).

$$\mathcal{A} - \mathcal{D} \le \varepsilon_\Delta \quad (33)$$

### B. Approximation and relaxation of bilinear components

Since there are a large number of bilinear products that affect the computing performance, specifically the generation cost $c_{it}^{\mathrm{g}} P_{it}^{\mathrm{g}}$ in (5) and (33) and the revenue $\tau_{it}^{\mathrm{p}} P_{it}^{\mathrm{g}}$ in (1), we applied relaxation techniques to handle these non-convex terms.

Firstly, we adopted the reformulated linearization technique (RLT) to initially relax the feasible domain of $c_{it}^g P_{it}^g$, namely $X_{it}^g$ hereafter, into a convex hull constructed by a set of linear constraints. As noted in [28], the two-dimensional RLT named the McCormick envelope is widely used in previous works.

$$\begin{aligned} X_{it}^g &\geq \underline{c}_{it} P_{it}^g, \\ X_{it}^g &\leq \overline{c}_{it} P_{it}^g, \\ X_{it}^g &\geq \overline{c}_{it} P_{it}^g + \overline{P}_{it}^g c_{it} - \overline{c}_{it} \overline{P}_{it}^g, \forall i \in \Psi_w \cup \Psi_v, \\ X_{it}^g &\leq \underline{c}_{it} P_{it}^g + \overline{P}_{it}^g c_{it} - \underline{c}_{it} \overline{P}_{it}^g, \forall t \in T. \end{aligned} \quad (34)$$

To narrow the above linear convex hull, the semi-definiteness condition is supplemented as (35). For convenience, we defined a series of vectors to contain the variable pair in bilinear products, detailed as $\mathbf{X}_{it} = [1, c_{it}, P_{it}^g]$, where 1 is an identity. It is noted that the element located at line 2 and row 3 of the symmetry matrix $(\mathbf{X}_{it})^T \mathbf{X}_{it}$ obtained by the cross product of the vector is $X_{it}^g$. As elaborated in [29], the relaxation derived from the combination of these methods can approximate the low-dimensional products accurately.

$$\begin{aligned} \mathbf{X}_{it} &\succcurlyeq 0 \\ (\mathbf{X}_{it})^T \mathbf{X}_{it} &\succcurlyeq 0, \forall i \in \Psi_w \cup \Psi_v, t \in T. \end{aligned} \quad (35)$$

As for the revenue summation $\tau_{it}^p P_{it}^g$ contained in (1), we applied the following steps to process this part. According to the first-order optimality conditions of $P_{it}^g$, this bilinear term can be equivalently restated as the sum of Lagrange multipliers in (36). Given that the complementary slackness of $P_{it}^g$ in (26), some elements in (36) can be further eliminated or transformed into linear variables. By introducing auxiliary variables $\zeta_{it}^{pg}$ constrained by big M relaxations in (37)-(38) to replace the product $\varphi_{it}^{pg} z_{it}^g$, $\tau_{it}^p P_{it}^g$ can be transformed into the sum of continuous linear variables. Therefore, the proposed bi-level framework is eventually transformed into a single-layer MISDP that can be solved by off-the-shelf solvers efficiently.

$$\begin{aligned} \sum_{t \in T} \tau_{it}^p P_{it}^g &= \sum_{t \in T} (c_{it}^g P_{it}^g - \lambda_{it}^{pg} P_{it}^g - \varphi_{it}^{pg} P_{it}^g) \\ &= \sum_{t \in T} c_{it}^g P_{it}^g - \sum_{t \in T} \varphi_{it}^{pg} z_{it}^g \overline{P}_{it}^g \\ &= \sum_{t \in T} X_{it}^g - \sum_{t \in T} \zeta_{it}^{pg} \overline{P}_{it}^g \end{aligned} \quad (36)$$

$$-M \times (1 - z_{it}^g) \leq \zeta_{it}^{pg} - \varphi_{it}^{pg} \leq M \times (1 - z_{it}^g) \quad (37)$$

$$-M \times z_{it}^g \leq \zeta_{it}^{pg} \leq M \times z_{it}^g, \forall i \in \Psi_w \cup \Psi_v, t \in T. \quad (38)$$

## IV. CASE STUDY

A modified IEEE 33-bus test feeder [31] with a high penetration level of DERs is adopted to test the proposed method, where the DER retailers are marked in Fig. 2. The reference voltage and capacity levels are 12.66kV and 1MW. The solar radiation and wind speed data are derived from the NREL Database [32]. The case study is solved using MATLAB with MOSEK on a laptop with an i7-10170 processor and 16GB RAM, and the generic parameters are summarized in Table I.

It is worth noting that the hosting capacity of DERs in the test feeder is sufficient to supply the system power demand during most of the non-peak hours. Besides, shunt capacitors are uniformly equipped at the supplier feeder to support power delivery. In the absence of direct intervention by government incentives, the deregulated retail market involving external power sources and DER retailers will be competitive. Since the LMP at the substation is predefined as a time-varying curve, it is undoubted that the profitability of a sole retailer is correlated with both the variation of LMP and rivals' strategies. Specifically, under the centralized clearing pattern, the bidding strategy of the marginal units in each timeslot will be the core information the participants are concerned about when bidding.

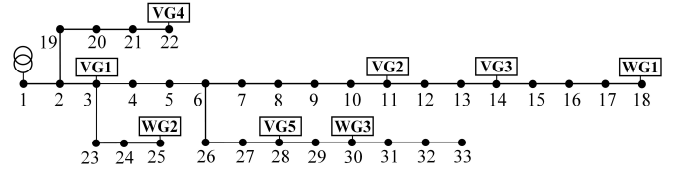

Fig. 2. IEEE 33-bus distribution network diagram separated by virtual agents

TABLE I
MAIN PARAMETERS AND SETTINGS IN CASE STUDY

| Name | Value | Name | Value | Name | Value |
|---|---|---|---|---|---|
| Capacity of WG | 1 MW | Capacity of VG | 0.5 MW | $\overline{P}_t^{tr}, \overline{Q}_{it}^{tr}$ | 5 MW |
| $\underline{v}_{it}, \overline{v}_{it}$ | 0.95, 1.05 | $R_i$ | $ 50 | $\theta_i$ | 3 $/MW |
| $\overline{Q}_{it}^c$ | 1 MW | $c^{tr}$ | 10 $/kg | $r^{gt}$ | 0.92 g/MW |
| $\overline{P}_{ij}$ | 3 MW | $\overline{Q}_{ij}$ | 3 MW | $\overline{L}_{ij}$ | 5 MA |
| $\varepsilon_\Delta$ | 1e0 | $t$ | 1 h | M | 1e3 |

To analyze the market performance of the proposed retail market mechanism under different conditions and the accuracy of the undirected power flow model, we designed several cases:
(1) Base case: Discussed in subsection A and B to provide a baseline of the proposed mechanism. The pricing strategy of the retailer is defined as time-invariant intervals.
(2) Traditional fixed pricing pattern: Discussed in subsection A to analyze the market efficiency by enacting the classic clearing mechanism in the proposed method with revenue constraints. The pricing strategy of retailers is set as fixed price curves.
(3) Mixed pricing interval case: Discussed in subsection B to reflect the market performance of a realistic market in the mechanism. The pricing strategy of the retailer is mixed with differentiated time-invariant and time-varying intervals.

### A. Base case

Under the premise of recovering life cycle investments, the pricing interval of retailers is set as time-constant in the base case. For simplicity, the intervals of wind and PV retailers are set as the same over the time horizon respectively, where $[\underline{c}_{it}^w, \overline{c}_{it}^w] = \$[20,30]$ and $[\underline{c}_{it}^v, \overline{c}_{it}^v] = \$[16,24]$. Also, the cumulative revenue thresholds of all retailers are set to be relatively small to avoid being naturally excluded from trading.

*1) Basic performance of market settlement*

The derived day-ahead DLMP is depicted in Fig. 3 as a set of statistical box plots with the variation of the LMP curve. It is observed from the green mini-graph enlarged at the bottom right that the nodal DLMP variates across the feeder at the same time, e.g. the 18:00. Also, this discrepancy varies between different hours. Since there are no constraint violations, the contemporaneous discrepancy of DLMP is caused by the growth of network losses from unit to user along the feeder.

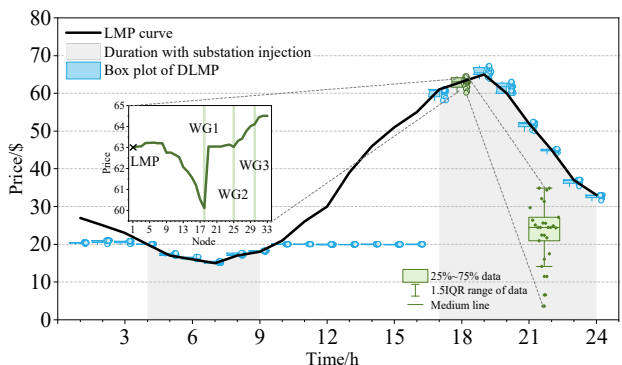

Fig. 3. Day-ahead DLMP of the entire system in base case

TABLE II
CUMULATIVE OUTPUT AND INCOME OF ENTITIES IN BASE CASE

| Entity | Output /MW | Rate[a] | Payoff /$ | Entity | Output /MW | Rate | Payoff /$ |
|---|---|---|---|---|---|---|---|
| WG1 | 9.51 | 0.46 | 395.04 | VG1 | 4.43 | 0.93 | 109.68 |
| WG2 | 7.49 | 0.55 | 268.07 | VG2 | 4.28 | 0.92 | 106.69 |
| WG3 | 8.98 | 0.66 | 300.45 | VG3 | 3.70 | 1.00 | 85.02 |
| EP[b] | 14.11 | / | 471.01 | VG4 | 4.28 | 0.92 | 106.62 |
| LA[c] | 60.00 | / | 1933.89 | VG5 | 3.70 | 1.00 | 85.64 |

[a]Rate represents the penetration rate; [b]EP denotes the external power plant that supplies the power injection through transmission feeder; [c]LA is the load aggregator of whole system.

Moreover, it is noticed from the grey area that the DLMP changes with the trend of the LMP only when the substation injection is nonzero. Theoretically, the clean energy with pricing intervals will be consumed at the top priority to enhance the social welfare defined in (5). But, if the lowest pricing $\underline{c}_{it}$ of available DERs is higher than the LMP, or the summation of renewable generations is insufficient to supply the demand, the substation injection will be dispatched to maintain secure operations. Also, it can be seen from the point data near each blue box that some DLMPs are cheaper than the LMP during peak hours, which shows the marginal unit includes not only the outer injection but also some economic DERs that operate almost at their limits. Taking the DLMP at 18:00 as an example, it is observed from the left mini-image of Fig. 3 that the DLMP at the wind unit locations has aroused significant price valleys.

To dissect the behavior of stakeholders, detailed cumulative generation outputs and payoffs are collected in Table II. Also, the final pricing and output curves of PV retailer 1 and wind retailer 2, abbreviated as VG1 and WG2 hereafter, are depicted on the left and right sides of Fig. 4. It is observed from Table II that the output and revenue of wind retailers are generally larger than that of PV retailers, whereas their penetration rates are relatively lower. This is because wind turbines are usually located far away from consumers, resulting in higher network losses and voltage drops than other suppliers, which is verified by the generation curtailments during 10:00-16:00 in the left graph of Fig. 4. Besides, it can be seen from Fig. 4 that the final pricing bids are both on the boundary of intervals, mostly at the lowest borderlines. The reason is that when the binary revenue restriction takes effect, the cheaper pricing of retailers is much conducive to the amelioration of social welfare. Besides, the cause for the pricing spike of WG2 during the 6:00-9:00 is the unreasonable pricing, i.e. the interval is completely higher than the cost of LMP, which has excluded the wind retailer from the transaction at this stage. In other words, the price surge at this time will not affect the settlement, which validates the effectiveness of the mechanism to avoid malicious bidding.

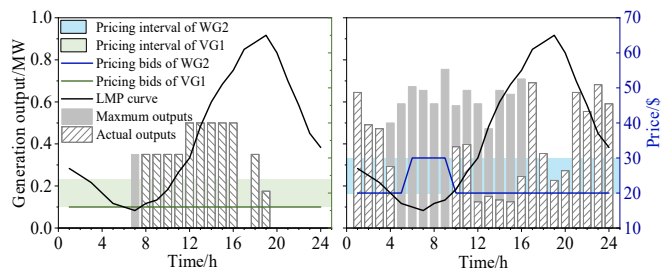

Fig. 4. Generation outputs and pricing bids of VG1 and WG2 in base case

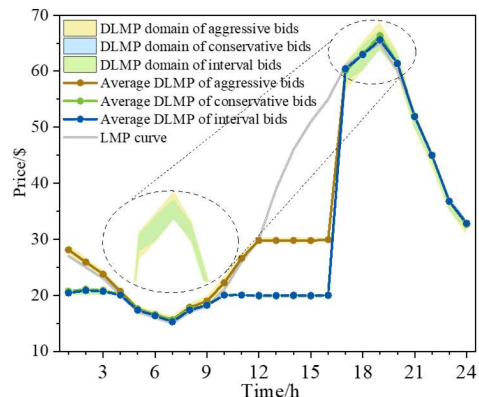

Fig. 5. DLMP domain and average DLMP curve of aggressive bid, conservative bid, and interval pricing bid

*2) Market efficiency of interval bidding*

To investigate the market efficiency, we compared the results derived from the traditional fixed pricing pattern with revenue considerations and the proposed mechanism. For simplicity, the pricing strategies considered in the traditional mode include conservative and aggressive proposals, which seek profit by enlarging the settlement quantity through lowing the unit price to the minimum boundary, and by rising clearing prices through increasing pricing to the maximum boundary, respectively.

It is observed from Fig. 5 that the DLMP curve obtained by the aggressive bid is relatively higher than other strategies in midnight and mid-day periods. As for the period between 1:00-4:00, as the LMP is cheaper than the price of available wind units, the power injection is dispatched as the sole supply source of the entire system, making the DLMP higher than LMP. During the period from 12:00 to 16:00, as the unit cost of wind retailers is reduced to a level lower than corresponding LMP, these wind units are served as marginal units when PV retailers are already operating at their limits, resulting in price cuts of DLMP. Besides, it is observed from the dashed circle that the DLMP derived from the interval bidding coincides with that obtained from the conservative pattern. The justification is that the LMP at this stage is always cheaper than the minimum boundary of pricing intervals, making the price reduction of retailers unable to promote additional power consumption.

However, if the pricing interval of clustered DERs is wide enough to envelop the LMP curve, it may be practical to improve the cumulative generation of DERs by applying the conservative mode, while the result is restricted by the revenue threshold. It is worth mentioning that higher settlement volume doesn't necessarily bring in more revenues correspondingly, which indicates the need to introduce pricing flexibility of zero-cost DERs to achieve capital recovery and gains. In essence, the binary revenue constraints increase the barrier for the DER retailer to enter the retail market, while the submitted parameter set, in turn, helps to reflect the discrepancies among participants and further establish a limited competitive retail market.

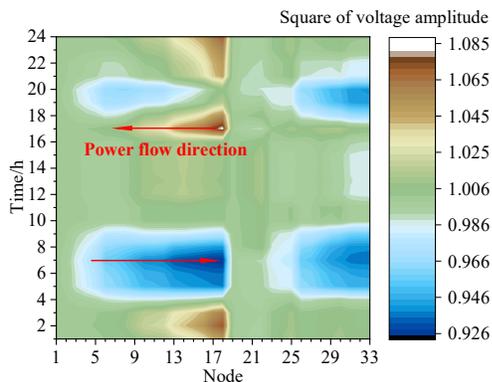

Fig. 6. Nodal voltage variation during a day of the test feeder in base case

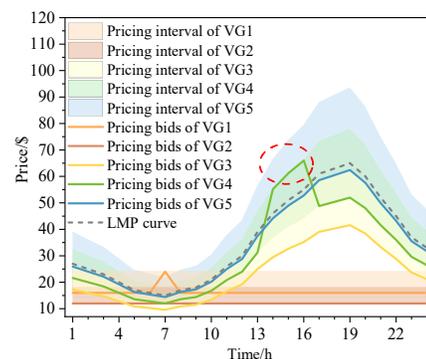

Fig. 7. Pricing bids and generation output of photovoltaic panels

TABLE III
CUMULATIVE OUTPUT AND INCOME OF ENTITIES IN MIXED BIDDING CASE

| Entity | Output /MW | Rate | Payoff /$ | Entity | Output /MW | Rate | Payoff /$ |
|---|---|---|---|---|---|---|---|
| WG1 | 14.76 | 0.71 | 560.21 | VG1 | 4.43 | 0.93 | 155.69 |
| WG2 | 13.69 | 1.00 | 444.29 | VG2 | 4.62 | 1.00 | 149.35 |
| WG3 | 8.27 | 0.60 | 288.32 | VG3 | 3.61 | 0.98 | 122.28 |
| EP | 7.35 | / | 360.08 | VG4 | 2.41 | 0.52 | 69.84 |
| LA | 60.00 | / | 2218.35 | VG5 | 1.53 | 0.41 | 54.59 |

*3) Exactness of undirected AC power flow model*

To analyze the accuracy of the solution derived from the undirected SOC-based power flow model, the daily node voltage distribution in the base case is depicted in Fig. 6. It is observed that the squared nodal voltage magnitude through the main feeder alters frequently, consequently indicating the power flow path changes with time-varying pricing. For the end bus 18, i.e. WG1, its voltage amplitude is higher than adjacent buses most of the time. Whilst during the 5:00-9:00 phase, this node occurs voltage drops, implying it evolves from a supplier to a consumer. Moreover, it is calculated that the optimal solution derived from the proposed undirected power flow model attains equality in the cone relaxations of (13)-(14) with a numerical precision of 1e-6 over the time horizon, which validates the numerical exactness of the proposed undirected SOC-based AC radial distribution power flow formulation.

*B. Competition among retailers with differentiated strategies*

The final market settlement scheme is closely related to the LMP curve which is undisclosed for all retailers. However, owing to incomplete information, each retailer may form a differentiated bidding strategy based on their estimates of LMP. To examine market performance under this circumstance, we redesigned the bidding strategy of retailers as a combination of time-constant and time-varying intervals. The pricing interval $[\underline{c}_{it}^{w}, \overline{c}_{it}^{w}]$ for three wind retailers are set as $[20, 30]$, $[0.8*\text{LMP}, 0.96*\text{LMP}]$, $[0.96*\text{LMP}, 1.44*\text{LMP}]$, and the pricing interval $[\underline{c}_{it}^{v}, \overline{c}_{it}^{v}]$ of PV retailers are $[16, 24]$, $[12, 18]$, $[0.8*\text{LMP}, 0.96*\text{LMP}]$, $[\text{LMP}, 1.2*\text{LMP}]$ and $[0.96*\text{LMP}, 1.44*\text{LMP}]$.

To explore the impact of binary revenue restrictions on the settlement, the final bidding schemes of PV retailers are collected in Fig. 7, and the related market info is listed in Table III. It is observed from Fig. 7 that the pricing bids of these retailers are mostly obtained at the lower border of submitted pricing intervals, except for the price soaring of VG1 and VG4. As the pricing interval of VG1 is the same as the base case, the reason lies in inappropriate pricing intervals. Nevertheless, as the interval of VG4 has already covered the LMP curve, the abnormal price spike denoted by the red circle may be affected by the binary revenue constraints. Although VG4 and VG5 are uneconomic among rivals, their lowest prices are cheaper than or equal to LMP, indicating their participation helps to improve social welfare when the revenue constraints take effect. Thus, the reason for the skyrocketing price of VG4 is to increase the output of VG5 by sacrificing its profit in an inactive way.

It is observed from Table III that the cumulative output of the two retailers is greatly reduced than that in the base case, and the revenue of VG5 eventually decreases to a level that closes to the revenue threshold. It reflects the fact that the proposed market mechanism does involve supervision to a certain extent, but such interventions that infringe on personal interests to improve overall social welfare are hindered by the submitted bidding information. The DSO-dominated decision will balance the social welfare enhanced by the retailer and its requirement. Generally, if a retailer submits both an excessively high revenue threshold and pricing intervals much higher than LMP, it will be automatically excluded from transactions.

*C. Discussion on the elimination of market power*

Since the proposed market mechanism is designed for a limited competitive retail market, it is necessary to discuss and eliminate market forces that may harm market order. Firstly, the predefined LMP is the spinning reserve cost of the network to limit the maximum cost of the marginal unit. Secondly, as LMP is formed based on the real-time operation of the bulk system, it is theoretically impossible to establish a conspiracy among all involved entities, which avoids the price manipulation based on LMP estimates. Besides, DSO can formulate some policies to prevent strategic mergers of stakeholders. Since the fairness index in (5) rewards retailers who submit economic bids, its weight can be increased to affect the preference of dispatching scheme. Also, DSO can adopt preventive rules to put penalties on strategic players. For instance, the repeatability of pricing intervals can be forbidden to avoid similar strategies.

## D. Error Analysis of approximations and relaxations

Since a series of approximation and relaxation techniques are applied in the pricing bidding process to estimate the final settlement scheme, it is necessary to quantify the error between the proposed method and the traditional economic settlement formulation where the obtained pricing bids are predefined. It is calculated that there is a slight positive/negative difference error between the DLMPs obtained by these different methods. However, the average absolute error of these results is less than 1% of the traditional solution, which illustrates the proposed mechanism can achieve acceptable accuracy in the settlement.

## V. CONCLUSIONS

A transactive retail market mechanism is presented to make coordinated decisions between the decoupled interval pricing from DER retailers and the economic market settlement from DSO simultaneously. In contrast to the involvement of direct game among entities to improve market efficiency, the profitability of DER retailers is modeled as a series of binary revenue constraints at the upper-level model to form pricing bids with income considerations, thus affecting the lower-level clearing. A bi-level virtual interplay among entities is executed by DSO to balance the profitability of retailers and improve clean energy penetration while maintaining market fairness. Owing to the uncertain power flow orientation, we improved the classic SOC-based AC power flow model to make it adaptable for undirected scenarios and also presented mild sufficient conditions to ensure the exactness of the solution. The next step is to develop an interactive market mechanism that fits for integrated energy systems with non-homogeneous players.

## APPENDIX A
### SUFFICIENT CONDITIONS FOR THE EXACTNESS OF UNDIRECTED SOC-BASED FORMULATION

Due to the reformulation of quadratic hyperplane into a convex cone, the classic directed second-order cone-based AC power flow model (DSOCM) essentially provides a lower bound to the economic dispatch of radial distribution circuits. The diagram of the feasible domain before and after cone relaxation is depicted in Fig.8. It is observed that the involvement of cone relaxation has enlarged the dark grey original hyperplane into an orange Lorenz cone, which then forms an intersection with the security operating bounding box, i.e. the polyhedron relaxed feasible domain in light grey.

As illustrated in [22]-[23], the exactness of the optimal solution derived from DSOCM is held under a series of strict sufficient conditions. However, these theoretical conditions may not apply to realistic networks [21]-[22]. Moreover, the research in [33] also pointed out the inherent obstacle of DSOCM in hardly describing the uncertain power flow scenarios for the future distribution system integrated with high rates of DERs.

To overcome these barriers, we improved the classic DSOCM to form an undirected SOC-based power flow model (USOCM) where the binary restriction is introduced to relax the rigorous orientation requirement. Also, we presented a mild sufficient condition to ensure the exactness of this model under some assumptions. In contrast to the latest research regarding the undirected formulation in [34], the shunt capacity is not considered in USOCM, while the over-theoretical limitation on infinite bus injection is relaxed.

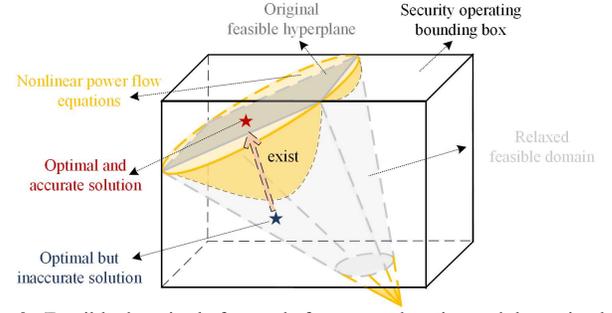

Fig. 8. Feasible domains before and after cone relaxation and the optimal solution searching in the proof

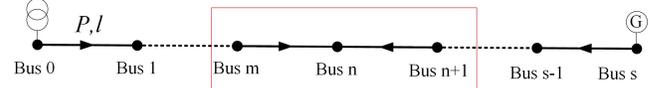

Fig. 9. Single distribution feeder with two-sided power sources

Given a tree structure of the radial distribution network, we use a single radial distribution feeder depicted in Fig.9 to elaborate the proof. By convention, bus 0 is the slack bus where the substation serves as a constant voltage source. Also, a large-capacity DER unit is accessed at the end of this feeder to support possible reverse power in the economic dispatch scheme. It is noted that the active power flow and line current orientations cannot be set owing to the intermittency and time-varying pricing of the renewable energy, which makes it necessary to form the power flow by using USOCM.

*Theorem:* Assume the objective function of the optimal USOCM is strictly convex, while the solution for power flow and nodal voltage is not binding. Then the optimal solution of USOCM will be exact if the below condition is ensured over the time horizon,

$$\breve{A}_{1,0} \cdots \breve{A}_{n,m} \breve{A}_{n+1,n} \cdots \breve{A}_{s,s-1} \begin{pmatrix} r_{s,s-1} \\ x_{s,s-1} \end{pmatrix} > 0 \quad (A.1)$$

Among them, $\breve{A}_{ij}$ is a constant related to parameter settings of the radial topology and secure boundary, detailed as below.

$$\breve{A}_{ij} = I - \frac{2}{\underline{v}_i} \begin{pmatrix} r_{ij} \\ x_{ij} \end{pmatrix} \left( [P_{ij}(\bar{p})]^+ \; [Q_{ij}(\bar{q})]^+ \right) \quad (A.2)$$

I is a $2 \times 2$ identity matrix. The calculator $[a]^+$ firmly restricts the reverse power flow $a$ to be greater than or equal to 0. $\bar{p}, \bar{q}$ denote the upper bounds of the bus net injection, which are negative for a consumer bus.

*Proof:* The above sufficient condition is firstly given in [35] to restrain the exactness of DSOCM for a reverse power flow scenario. Following [35], we will explain from the portability and practicality perspectives that this sufficient condition is also suitable for the practical application of USOCM.

In terms of portability, a linear Dist-flow model that neglects line current terms is utilized to approximate the worst power flow solution. It is noted that when the source at end bus s serves as the sole supplier, the solution derived from this linear model will provide an upper bound for the state variables of the single feeder in Fig.9. As can be seen from the red block in Fig.9, compared with the bi-source situation modeled by USOCM where the position of the power distribution point is unclear, the complete reverse flow case in the Dist-flow model makes the end node voltage rising to the top and the power flow near the source increasing.

If it is assumed the optimal solution obtained from USOCM cannot be attained equality in cone relaxations, following the recursive Dist-flow calculations in [35], we can always find another superior solution that contradicts the optimality in this hypothesis. It can be seen from Fig. 8 that this process is to find another feasible point that requires less cost than the navy presumed optimal solution, and then it can be illustrated that the newly discovered red solution is actually the optimal solution with accuracy property.

As the state variables in the optimal solution are not binding and even far away from their secure limits, it can be ensured the variables in the superior solution obtained from the Dist-flow model are feasible under safe operating conditions. Therefore, it implies the optimality and the exactness of the solution obtained from USOCM can be maintained concurrently in a single distribution feeder with two-sided power sources. As noted in remark 1 of [35], the involvement of other suppliers will retain the exactness result, which broadens the applicability of USOCM to a radial distribution network with high-level DER integration.

As for the practicality, it is generally presumed the objective function of DSOCM to be convex [22]-[23], [34]. Specifically, it requires this goal to be strictly increasing with the growth of bus net injection $(p_i, q_i)$ along with squared line current $l_{ij}$, and also independent of the variation of active and reactive power $P_{ij}, Q_{ij}$. As the minimization of generation cost is the main concern of most economic dispatching, this assumption is naturally held.

Moreover, since the impedance of the distribution line is generally far less than 1, whilst the power flow in a distribution network is smaller than the sole source case, the element in $\check{A}_{ij}$ will always be approaching or even equal to 1 in reality, thus making the condition in (A.1) to be practically satisfied in most realistic cases. Besides, as the voltage regulator and the operating capacity of distribution topology are sufficient in the active distribution network, there is less possible to lead to state variables operating on its boundaries in the optimal solution. It reveals the universality of USOCM to obtain exact solutions in realistic scenarios.

## APPENDIX B
### DUAL COUNTERPART OF THE LOWER-LEVEL SOCP

The equivalent dual model of the relaxed lower-level SOCP is formulated as below. The first-order conditions of (B.7a), (B.8a), and (B.9a) are the derivatives of augmented Lagrange function for the forward parameters $P_{ijt}^+, Q_{ijt}^+, l_{ijt}^+$, and the rest in pairs are derived from the reverse parameters $P_{ijt}^-, Q_{ijt}^-, l_{ijt}^-$.

$$\max \mathcal{D}$$

$$= \sum_{t \in T} \begin{bmatrix} \sum_{i \in \Psi_N}(P_{it}^L \tau_{it}^p + Q_{it}^L \tau_{it}^q) \\ + \sum_{i \in \Psi_{\{1\}}}(\bar{P}_{it}^g \varphi_{it}^{pg} + \bar{Q}_{it}^g \varphi_{it}^{qg}) \\ + \sum_{i \in \Psi_{N \setminus \{1\}}}(\underline{v}_{it} \lambda_{it}^v + \bar{v}_{it} \varphi_{it}^v) + \omega_t^{bn} \\ + \sum_{i \in \Psi_c} \bar{Q}_{it}^c \varphi_{it}^c + \sum_{i \in \Psi_w \cup \Psi_v} \varphi_{it}^u \\ + \sum_{ij \in \Psi_f}(\varphi_{it}^{z+} + \varphi_{it}^{z-} + \omega_{it}^z) \end{bmatrix} \quad (B.1)$$

$\text{subject to.} \quad \tau_{it}^p + \lambda_{it}^{pg} + \varphi_{it}^{pg} = \text{LMP}_t, \forall i \in \Psi_{\{1\}}, \quad (B.2)$

$\tau_{it}^p + \lambda_{it}^{pg} + \varphi_{it}^{pg} = c_{it}^g, \forall i \in \Psi_w \cup \Psi_v, \quad (B.3)$

$\tau_{it}^q + \lambda_{it}^{qg} + \varphi_{it}^{qg} = 0, \forall i \in \Psi_{\{1\}} \cup \Psi_w \cup \Psi_v \cup \Psi_c, \quad (B.4)$

$\sum_{j \in \Omega_i^+}(\omega_{ijt} + d_{ijt}^+ + d_{ijt}^- - d_{ijt}^{C+} - d_{ijt}^{C-})$
$\quad - \sum_{k \in \Omega_i^-} \omega_{kit} + \omega_t^{bn} = 0, \forall i \in \Psi_{\{1\}}, \quad (B.5)$

$\sum_{j \in \Omega_i^+}(\omega_{ijt} + d_{ijt}^+ + d_{ijt}^- - d_{ijt}^{C+} - d_{ijt}^{C-})$
$\quad - \sum_{k \in \Omega_i^-} \omega_{kit} + \lambda_{it}^v + \varphi_{it}^v = 0, \forall i \in \Psi_{N \setminus \{1\}}, \quad (B.6)$

$-\tau_{it}^p + \tau_{jt}^p - 2r_{ij}\omega_{ijt} + d_{ijt}^{A+} + \lambda_{ijt}^{p+} + \varphi_{ijt}^{p+} = 0, \quad (B.7a)$

$-\tau_{it}^p + \tau_{jt}^p - 2r_{ij}\omega_{ijt} - 2d_{ijt}^{A-} + \lambda_{ijt}^{p-} + \varphi_{ijt}^{p-} = 0, \quad (B.7b)$

$-\tau_{it}^q + \tau_{jt}^q - 2x_{ij}\omega_{ijt} + d_{ijt}^{B+} + \lambda_{ijt}^{q+} + \varphi_{ijt}^{q+} = 0, \quad (B.8a)$

$-\tau_{it}^q + \tau_{jt}^q - 2x_{ij}\omega_{ijt} - 2d_{ijt}^{B-} + \lambda_{ijt}^{q-} + \varphi_{ijt}^{q-} = 0, \quad (B.8b)$

$(r_{ij}^2 + x_{ij}^2) \cdot \omega_{ijt} + \lambda_{ijt}^{f+} + \varphi_{ijt}^{f+}$
$\quad + d_{ijt}^+ + d_{ijt}^{C+} - r_{ij}\tau_{it}^p - x_{ij}\tau_{it}^q = 0, \quad (B.9a)$

$(r_{ij}^2 + x_{ij}^2) \cdot \omega_{ijt} + \lambda_{ijt}^{f-} + \varphi_{ijt}^{f-}$
$\quad - d_{ijt}^- - d_{ijt}^{C-} + r_{ij}\tau_{it}^p + x_{ij}\tau_{it}^q = 0, \quad (B.9b)$

$\left| d_{ijt}^+ ; d_{ijt}^{A+} ; d_{ijt}^{B+} ; d_{ijt}^{C+} \right| \preccurlyeq 0, \quad (B.10)$

$\left| d_{ijt}^- ; d_{ijt}^{A-} ; d_{ijt}^{B-} ; d_{ijt}^{C-} \right| \preccurlyeq 0, \quad (B.11)$

$-(\bar{P}_{ij}\varphi_{ijt}^{p+} + \bar{Q}_{ij}\varphi_{ijt}^{q+} - \bar{Q}_{ij}\lambda_{ijt}^{q+} + \bar{L}_{ij}\varphi_{ijt}^{f+}) + \lambda_{it}^{z+}$
$\quad + \varphi_{it}^{z+} + \omega_{it}^z = 0, \quad (B.12)$

$\bar{P}_{ij}\varphi_{ijt}^{p-} + \bar{Q}_{ij}\varphi_{ijt}^{q-} - \bar{Q}_{ij}\lambda_{ijt}^{q-} + \bar{L}_{ij}\varphi_{ijt}^{f-} + \lambda_{it}^{z-}$
$\quad + \varphi_{it}^{z-} + \omega_{it}^z = 0, \forall ij \in \Psi_f, \quad (B.13)$

$-(\bar{P}_{it}^g\varphi_{it}^{pg} + \bar{Q}_{it}^g\varphi_{it}^{qg}) + \lambda_{it}^u + \varphi_{it}^u = 0,$
$\quad \forall i \in \Psi_w \cup \Psi_v, t \in T. \quad (B.14)$

## APPENDIX C
### STRONG DUALITY RELAXATION FOR SOCP

In contrast to several existing AC power flows where network losses and voltage offsets are approximated by linear functions, the result attained by the classic SOC-based power flow model is much accurate and even exact under specific conditions. For brevity, the mathematic formulation of a primal SOCP optimal power flow model is presented compactly.

$$\min \quad c^T x \quad (C.1)$$
$$\text{subject to} \quad Ax \geq b : \mu \quad (C.2)$$
$$\|u_i\| \leq t_i, i = 1, 2, \cdots, L \quad (C.3)$$
$$u_i = E_i x + f_i : z_i \quad (C.4)$$
$$t_i = G_i^T x + h_i : w_i \quad (C.5)$$

Among them, (C.2) is a linear constraint regarding the generation boundaries and the bounding box of state variables, and (C.3) is the second-order cone about the line capacity. $x \in$

$\mathbb{R}^n$ is the variable, and the coefficient matrixes are $c \in \mathbb{R}^n$, $b \in \mathbb{R}^m$, $A \in \mathbb{R}^{m \times n}$, $E_i \in \mathbb{R}^{(n_i-1) \times n}$, $f_i \in \mathbb{R}^{n_i-1}$, $G_i \in \mathbb{R}^n$, $h_i \in \mathbb{R}$.

Based on [36], the dual counterpart of SOCP is given below, where Lagrange multipliers are $\mu \in \mathbb{R}^m$, $z_i \in \mathbb{R}^{n_i-1}$, $w_i \in \mathbb{R}$, and (C.8) is the dual second-order cone constraint.

$$\max \quad b^T \mu - \sum\nolimits_{i=1}^{L} (f_i^T z_i + h_i w_i) \quad \text{(C.6)}$$

$$\text{subject to} \quad A^T \mu + \sum\nolimits_{i=1}^{L} (E_i^T z_i + G_i w_i) = c \quad \text{(C.7)}$$

$$\|z_i\| \leq w_i, i = 1, 2, \cdots, L \quad \text{(C.8)}$$

$$\mu \geq 0 \quad \text{(C.9)}$$

The Karush-Kuhn-Tucker (KKT) optimality condition of SOCP including complementary slackness conditions of (C.2) and (C.3) along with the first-order optimality regarding $x$ are listed below. It is well known to be a way to obtain primal and dual solutions concurrently. It is worth noting that since the SOC relaxation should attain equality to make the solution feasible for the original nonlinear power flow formulation, the slackness condition of (C.2) should be listed as (C.11).

$$\mu^T(Ax - b) = 0 \quad \text{(C.10)}$$

$$z_i^T u_i + w_i t_i = 0, i = 1, 2, \cdots, L \quad \text{(C.11)}$$

$$x^T \left[ c - A^T \mu - \sum\nolimits_{i=1}^{L} (E_i^T z_i + G_i w_i) \right] = 0 \quad \text{(C.12)}$$

$$\text{primal constraints: (C.2)} - \text{(C.5)} \quad \text{(C.13)}$$

$$\text{dual constraints: (C.7)} - \text{(C.9)} \quad \text{(C.14)}$$

However, since the above complementarity has introduced intractable compound products, we analyzed the possibility of relaxing the complementarity appropriately. Inspired by [27], a model aimed at controlling positive relaxation values of KKT conditions is reconstructed in the following to approximate the equivalent KKT condition of the SOCP.

$$\min \quad \varepsilon_1 + \varepsilon_2 + \varepsilon_3 \quad \text{(C.15)}$$

$$\text{subject to} \quad \text{(C.2)} - \text{(C.5)}, \text{(C.7)} - \text{(C.9)} \quad \text{(C.16)}$$

$$\mu^T(Ax - b) \leq \varepsilon_1, \quad \text{(C.17)}$$

$$z_i^T u_i + w_i t_i \leq \varepsilon_2, \quad \text{(C.18)}$$

$$x^T \left[ c - A^T \mu - \sum\nolimits_{i=1}^{L} (E_i^T z_i + G_i w_i) \right] \leq \varepsilon_3, \quad \text{(C.19)}$$

$$\varepsilon_1, \varepsilon_2, \varepsilon_3 \geq 0$$

Theoretically, the above goal in (C.13) combined with relaxed conditions in (C.15)-(C.17) can be transformed equivalently as shown below.

$$\min \quad \varepsilon_1 + \varepsilon_2 + \varepsilon_3$$
$$\Leftrightarrow$$
$$\min \quad \begin{aligned} & \mu^T(Ax-b) + \sum\nolimits_{i=1}^{L}(z_i^T u_i + w_i t_i) \\ & + x^T \left[ c - A^T\mu - \sum\nolimits_{i=1}^{L}(E_i^T z_i + G_i w_i) \right] \end{aligned}$$
$$= c^T x - b^T \mu - \sum\nolimits_{i=1}^{L} (z_i^T E_i x + w_i G_i^T x) \quad \text{(C.20)}$$
$$\quad + \sum\nolimits_{i=1}^{L} (z_i^T u_i + w_i t_i)$$
$$= c^T x - b^T \mu - \sum\nolimits_{i=1}^{L} [z_i^T(u_i - f_i)$$
$$\quad + w_i(t_i - h_i) + z_i^T u_i + w_i t_i]$$
$$= c^T x - \left[ b^T \mu - \sum\nolimits_{i=1}^{L} (f_i^T z_i + h_i w_i) \right]$$

It can be seen that the minimization of the restriction regarding relaxation degree is essentially equivalent to decrease the duality gap between the primal and dual models of SOCP. Therefore, the KKT optimality conditions of SOCP can be ultimately approximated by a model that targets to minimize its duality gap.

$$\min \quad c^T x - \left[ b^T \mu - \sum\nolimits_{i=1}^{L} (f_i^T z_i + h_i w_i) \right] \quad \text{(C.21)}$$

$$\text{subject to. primal constraints: (C.2)} - \text{(C.5)} \quad \text{(C.22)}$$

$$\text{dual constraints: (C.7)} - \text{(C.9)} \quad \text{(C.23)}$$

If the value of the objective function in (C.21) is to be or near to be zero, it means the result derived from this single-layer minimum model are consistent or closes to the solution obtained by solving the primal and dual models sequentially.

ACKNOWLEDGMENT

The authors wish to acknowledge Professor Jianhui Wang, from Southern Methodist University, for his help in advising the formulation of the transactive framework of this study.